\documentclass[11pt,a4paper]{article}
\usepackage{amssymb,amsfonts,amsmath,amsthm,euscript}
\usepackage[english]{babel}
\usepackage[utf8]{inputenc}
\usepackage{amsfonts, dsfont}
\usepackage{graphicx}
\usepackage{subfigure}
\usepackage{color}
\usepackage{mathrsfs}
\usepackage{enumerate,graphics}
\usepackage{multicol}
\usepackage{multirow}
\usepackage{lscape}
\usepackage{placeins}
\usepackage{natbib}
\usepackage{float}
\usepackage[T1]{fontenc}

\usepackage[implicit=false]{hyperref}
\usepackage[left=2.7cm,right=2.7cm,bottom=2.67cm,top=2.67cm]{geometry}
\allowdisplaybreaks[4]

\newcommand{\Z}{\mathds Z} 
\newcommand{\R}{\mathds R} 
\newcommand{\E}{\mathds E} 
\newcommand{\C}{\mathds C}
\newcommand{\Ca}{\mathcal C^\ast}

\newcommand{\eps}{\varepsilon}

\newcommand{\bs}[1]{\boldsymbol{#1}}

\newcommand{\F}{\mathscr{F}}
\newcommand{\var}{\mathop{\mathrm{Var}}}

\newcommand{\ti}[1]{\mbox{\tiny\ensuremath{#1}}}
   
\newcommand{\pumi}[1]{{\color{red}{#1}}}        
\newcommand{\pumis}[1]{{\expandafter \xout \expandafter{#1}}}

\long\def\sfootnote[#1]#2{\begingroup%
\def\thefootnote{\fnsymbol{footnote}}\footnote[#1]{#2}\endgroup}
\def\bfootnote{\xdef\@thefnmark{}\@footnotetext}

\begin{document}
\thispagestyle{empty}
{
\centering \Large{\bf Bayesian Analysis of Beta Autoregressive Moving \\ Average Models}\vspace{.5cm}\\
\normalsize{ {\bf Aline Foerster Grande$\!\phantom{i}^{\mathrm{a}}$\let\thefootnote\relax\footnote{\hskip-.3cm$\phantom{s}^\mathrm{a,}$Programa de P\'os Gradua\c c\~ao em Estat\'istica - Universidade Federal do Rio Grande do Sul.
}, Guilherme Pumi$\!\!\phantom{s}^\mathrm{a,}$\sfootnote[1]{Corresponding author. This Version: \today.} and Gabriela Bettella Cybis${}^\mathrm{a}$
 \\
\let\thefootnote\relax\footnote{E-mails: alinefoerstergrande@gmail.com (A.F. Grande); guilherme.pumi@ufrgs.br (G. Pumi); gcybis@gmail.com (G.B. Cybis)}}\\
\let\thefootnote\relax\footnote{ORCIDs: 0000-0003-4535-9909 (Grande); 0000-0002-6256-3170 (Pumi) and 0000-0002-2791-6735 (Cybis).}\\
\vskip.3cm
}}

\begin{abstract} 
This work presents a Bayesian approach for the estimation of Beta Autoregressive Moving Average ($\beta$ARMA) models. We discuss standard choice for the prior distributions and employ a Hamiltonian Monte Carlo algorithm to sample from the posterior. We propose a method to approach the problem of unit roots in the model's systematic component. We then present a series of Monte Carlo simulations to evaluate the performance of this Bayesian approach. In addition to parameter estimation, we evaluate the proposed approach to verify the presence of unit roots in the model's systematic component and study prior sensitivity. 
An empirical application is presented to exemplify the usefulness of the method. In the application, we compare the fitted Bayesian and frequentist approaches in terms of their out-of-sample forecasting capabilities. 
\end{abstract}
\linespread{1.1}
\section{Introduction}   

The literature related to regression models for time series has grown rapidly in the last few years. A particularly fertile area of research involves the theory of bounded time series regression. Typical examples are ratios and proportions observed over time. In these situations, it is well-known that ARMA models are not adequate. One of the most employed methods to deal with bounded time series is based on the GARMA (Generalized ARMA) approach, introduced under this name by \cite{benjamin2003} \citep[see][for earlier approaches]{Zeger, Li1994}. The idea is to combine the flexibility of the generalized linear model (GLM) structure with the classical ARMA modeling approach. The GLM structure addresses the problem of bounds while still allowing for the presence of covariates, used to accommodate different structures, such as heteroscedasticity, trends, etc.; whereas the time series structure accommodates the presence of serial correlation in the process. Although simple, the approach is quite general. 

After \cite{benjamin2003}, the GARMA structure has evolved in various forms, by assuming distributions outside the exponential family \citep{Rocha2009,bayers,RARMA,PTSR,UWARMA}, often referred to as GARMA-like models, including long-range dependence \citep{PUMI2019} and even chaotic dependence structure \citep{BARC}. 
GARMA and GARMA-like models are typically formulated as observation-driven models, and inference is often based on conditional or partial maximum likelihood  \citep[][among others]{Rocha2009,bayers,PTSR,UWARMA,PUMI2019,BARC}, although early models considered a quasi-maximum likelihood approach \citep{Zeger, Li1994}. In the context of $\beta$ARMA models, parameter estimation using conditional and partial maximum likelihood has been studied in \cite{Rocha2009} and \cite{PUMI2019}. These methods are implemented in \texttt{R} \citep{R} package  \texttt{BTSR} \citep{BTSR}. Additionally, small-sample corrections for conditional maximum likelihood can be found in \cite{Bruna}.

Regression models for time series can be categorized into two types, as per the nomenclature of \cite{cox1981}: parameter-driven and observation-driven models. In the first type autocorrelation is introduced through a latent variable.
In the parameter-driven context, dynamical generalized linear models (DGLM) exhibit a similar structure to GARMA models.  Bayesian estimation for DGLM has a long-standing literature, commencing with the seminal work of \cite{west1985} for dynamic generalized linear models (DGLM) based on the exponential family. MCMC methods for the Bayesian analysis of DGLM models were developed by \cite{daniB}.  In a multivariate setup, \cite{GRG} proposed a similar approach to \cite{west1985} to model multivariate time series of continuous proportions. \cite{godo} considered the decomposition of DGLM models into sums of noise-free simpler dynamic linear models. \cite{souza} recently extended DGLM to cover the two-parameter exponential family. 

In the context of observation-driven GARMA and GARMA-like models, Bayesian estimation has received comparatively less attention. 
Bayesian estimation of GARMA models for count time series is studied in \cite{bgarma} while \cite{casarin} considers a Bayesian approach to model selection in $\beta$AR models. \cite{cepeda} considers Bayesian estimation in the context of $\beta$ regression proposed by \cite{Ferrari2004}. For a comprehensive review and comparison between observation-driven and parameter-driven models in the context of GARMA models for count time series, we refer to the work of \cite{rev} and the references cited therein.

Considering the Bayesian estimation of $\beta$ARMA models, the only specific paper available is \cite{casarin}, where the authors consider a Bayesian approach to model selection in $\beta$AR models, without covariates, using a reversible jump Markov chain approach. The framework considered in the paper, however, is quite restrictive because the authors propose to use a linear link in the model's systematic component, requiring the imposition of conditions on the model's coefficients to keep the conditional mean within the interval $(0,1)$. In addition, because it is non-standard, there are no statistical packages available to fit such a particular model, which hinders its application.

In this paper, we propose the study of a fully specified $\beta$ARMA model under the Bayesian framework. The model considers standard forms for the prior distributions, for which we make uninformative parameter choices.  We employ the Hamiltonian Markov Chain Monte Carlo (MCMC) method to obtain the posterior distribution of the model parameters, and the No-U-Turn sampler (NUTS) for Hamiltonian parameter fine-tuning. We also use the model to assess the posterior probability that the model's systematic component contains a quasi-unit root. Furthermore, we consider a procedure consisting of a series of Bayes Factor (BF) tests to determine the order $p$ and $q$ of the $\beta$ARMA$(p,q)$ model, and address the issue of prediction by assessing the posterior predictive distribution. The quality of parameter estimates and sensibility to prior specifications is assessed through simulation studies.  The proposed framework allows for a comprehensive fully Bayesian analysis of bounded time series data under the $\beta$ARMA$(p,q)$ model, complete with parameter estimation, sensitivity analysis, identification of unit roots, and prediction.  

We also present a Monte Carlo simulation study to verify the performance of the proposed approach. We evaluate point and interval estimation under various scenarios in the simulation, approach the problem of unit roots in $\beta$ARMA's systematic component, and provide a prior sensitivity analysis. We showcase the proposed methodology in an empirical application to hydroelectric energy storage. We perform model selection based on the proposed Bayes factor approach, comparing the selected model's out-of-sample forecasting capabilities with competitor models, including a model selected using the frequentist approach. 

The following is a summary of the paper. Section \ref{BARMA_Model} comprehensively reviews the class of $\beta$ARMA models. Section \ref{Bayesian_section} delves into key concepts associated with the Bayesian inference, such as the Hamiltonian Monte Carlo algorithm, posterior predictive distribution, and quasi-unit root testing. In Section \ref{simulation}, we carry out a Monte Carlo simulation to assess the finite sample behavior of the proposed approach. Finally, Section \ref{aplication} provides an illustrative example to showcase the practical application of the proposed method.
\section{\texorpdfstring{$\beta$}{B}ARMA Models}   \label{BARMA_Model}
Let $\{Y_t\}_{t\in\Z}$ be a time series of interest satisfying $Y_t\in(0,1)$ for all $t\in\Z$, and let $\{\bs X_t\}_{t\in\Z}$ denote a set of $r$-dimensional exogenous time dependent (possibly random) covariates. Let $\F_{t}$ denote the $\sigma$-field representing the observed history of the model up to time $t$, that is, the sigma-field generated by $(\bs X_t^\prime,Y_t,\bs X_{t-1}^\prime,Y_{t-1},\cdots)$. The $\beta$ARMA$(p,q)$ model introduced by \cite{Rocha2009} is an observation-driven model in which the random component follows a conditional beta distribution, parameterized in terms of its mean and a precision parameter as
\begin{align}\label{e:density}
p(y|\nu,\mu_t,\F_{t-1})=\frac{\Gamma(\nu)}{\Gamma(\nu\mu_t)\Gamma\big(\nu(1-\mu_t)\big)}\,y^{\nu\mu_t-1}(1-y)^{\nu(1-\mu_t)-1},
\end{align}
for $0<y<1$, $\mu_t:=\E(Y_t|\F_{t-1})$ and $\nu>0$. Observe that  $\var(Y_t|\F_{t-1})=\frac{\mu_t(1-\mu_t)}{1+\nu}$, so that the model is conditionally heteroscedastic. Additionally, $\nu$ can be interpreted as a precision parameter in the sense that, the higher the $\nu$, the smaller the conditional variance. Moreover, $\var(Y_t|\F_{t-1})\leq \frac1{4\nu}\rightarrow 0$ as $\nu$ goes to infinity. Hence, in practice, very high values of $\nu$ account for (conditional) homoscedastic behavior.
Let $g:(0,1)\rightarrow\R$ be a twice-differentiable strictly monotonic link function. The most commonly applied link functions are the logit (applied here), probit, and complementary log-log, although parametric alternatives have been explored in the literature \citep{Pumi2020}. The systematic component in a $\beta$ARMA$(p,q)$ model follows an ARMA$(p,q)$-like structure prescribed by
\begin{align}\label{model}
g(\mu_t):= \alpha + \bs{X}_t' \bs\beta + \sum_{i=1}^p \phi_i \big( g(Y_{t-i})-\bs X_{t-i}' \bs\beta\big) + \sum_{j=1}^q \theta_j r_{t-j},
\end{align}
where $r_t:=g(Y_t)-g(\mu_t)$ denotes an error term,  $\bs\beta=(\beta_1,\cdots,\beta_r)'$ is the $r$-dimensional vector of parameters related to the covariates, $\bs\phi=(\phi_1,\cdots,\phi_p)'$ and $\bs\theta=(\theta_1,\cdots,\theta_q)'$ are the AR and MA coefficients, respectively. The $\beta$ARMA$(p,q)$ model is  defined by \eqref{e:density} and \eqref{model}. 

Parameter estimation in $\beta$ARMA models is usually conducted using a frequentist approach, based on the conditional \citep{Rocha2009, Rocha2009err} or partial \citep{PUMI2019} maximum likelihood, which is reflected in the notation here. 
In the next section we propose a fully Bayesian approach for parameter estimation in the context of $\beta$ARMA$(p,q)$ models.
\section{Bayesian inference} \label{Bayesian_section}
Let $y_1,\cdots,y_n$ be a sample from a $\beta$ARMA$(p,q)$ model, and $\bs{\gamma} := (\nu,\alpha,\bs \beta', \bs\phi', \bs\theta')'$. The partial likelihood function is given by 
\begin{equation*}
L(\bs\gamma|\F_{n})\propto\prod_{t=1}^n p(y_t;\mu_t,\nu|\F_{t-1}) \propto \prod_{t=1}^n\frac{y_t^{\nu\mu_t-1}(1-y_t)^{\nu(1-\mu_t)-1}}{\Gamma(\nu\mu_t)\Gamma\big(\nu(1-\mu_t)\big)}\,.
\end{equation*}
For the non negative precision parameter $\nu$ we consider a gamma prior of the form $\nu\sim \mathrm{Gamma}(a,b)$ (mean $a/b$). The link function in \eqref{model} guarantees that $\mu_t\in(0,1)$ for all $\alpha$, $\bs\beta\in\R^r$, $\bs\theta\in\R^q$ and $\bs\phi\in\R^p$, thus we shall  assume independent multivariate normal priors of the form 
\begin{equation}\label{mutla}
\alpha\sim N(0,\sigma_\alpha^2),\quad \bs\beta \sim N_r(\bs 0,\sigma_\beta^2I_r),\quad\bs\phi \sim N_q(\bs0,\sigma_\phi^2I_p),\quad \bs\theta\sim N_q(\bs0,\sigma_\theta^2I_q),
\end{equation}
where $N_s(\bs 0,K)$ denotes the $s$-variate  normal distribution with mean $\bs0\in\R^s$ and variance-covariance matrix $K$, while $I_s$ denotes the $s\times s$ identity matrix.

When no prior knowledge regarding the variance hyper-parameters $\sigma_\alpha^2,\sigma_\beta^2, \sigma_\phi^2$ and $\sigma_\theta^2$ is available, we shall consider very high values of these parameters so that the prior will be non-informative. As for the hyper-parameters related to $\nu$, in section \ref{sense} we shall conduct a sensitivity analysis regarding their choice. 

Combining the joint prior density, denoted by $\pi_0(\bs\gamma)$, with the likelihood function, we note that the posterior distribution is given by  
\begin{equation} \label{post}
\pi(\bs\gamma|\F_{n})\propto L(\bs\gamma|\F_{n})\pi_0(\bs\gamma).
\end{equation}
Unfortunately, the joint posterior density \eqref{post} cannot be obtained in closed form. To sample from the posterior distribution we adopt a Hamiltonian Monte Carlo scheme.
\subsection{Hamiltonian Monte Carlo (HMC)}
Hamiltonian Monte Carlo is a state-of-the-art general-purpose sampling technique based on the Hamiltonian dynamics concomitant with a Metropolis-Hastings transition rule \citep{Duane,Mackay,Neal}. It is generally an efficient alternative for simulating random values from a target distribution and requires the evaluation of the log density and its gradient. In order to sample from the joint distribution of  $\bs{\gamma}$, a momentum variable $\bs\kappa \sim N_d(\bs 0, I_d)$,   independent of $\bs\gamma$, is added to the dynamics and sampled along with $\bs\gamma$. In this case, the Hamiltonian is defined by 
\begin{equation*}
H(\bs\kappa,\bs\gamma):=\log\big(\pi(\bs\gamma|\F_{n})\big)+\frac12\bs\kappa'\bs\kappa,
\end{equation*}
and the sample space of $(\bs\kappa,\bs\gamma)$ is explored according to a Hamiltonian dynamics, namely,
\begin{align}\label{hd}
\frac{\partial H(\bs\kappa,\bs\gamma)}{\partial \bs\kappa}=\bs\kappa' \quad\mbox{and}\quad \frac{\partial H(\bs\kappa,\bs\gamma)}{\partial \bs \gamma} = \Delta_{\bs\gamma}\log\big(\pi(\bs\gamma|\F_{n})\big).
\end{align}
To approximate a solution of \eqref{hd} and obtain an iteration of the HMC, the first step is to sample $\bs\kappa$ from a $N_d(\bs 0, I_d)$. Next, dynamics \eqref{hd} is approximated from time $t=0$ to time $t=\tau$ by applying $\lfloor \tau/\eps\rfloor$ Leapfrog steps of size $\eps$, namely,
\begin{equation}\label{half}
\bs\kappa \leftarrow \bs\kappa + \frac\eps2 \Delta_{\bs\gamma}\log\big(\pi(\bs\gamma|\F_{n})\big), \quad \bs\gamma\leftarrow \bs\gamma + \bs\kappa\eps, \quad \bs\kappa \leftarrow \bs\kappa + \frac\eps2 \Delta_{\bs\gamma}\log\big(\pi(\bs\gamma|\F_{n})\big).
\end{equation}
It can be shown that \eqref{half} yields a valid Metropolis proposal that is accepted or rejected using standard acceptance probabilities. Finely tuned values of $\eps$ and $\tau$ are crucial for HMC's performance.
An automatic way to fine-tune the choice of these hyper-parameters is by using the No U-Turn Sampler (NUTS) algorithm, created by \cite{Hoffman}. We use package \texttt{RStan} \citep{rstan} in \texttt{R} \citep{R} to perform the HMC.
\subsection{Posterior predictive distribution}\label{pred}

Let $\tilde{\bs Y} = (Y_{n+1},\cdots, Y_{n+h})'$ represent the new observations for which we wish to make predictions. Then the posterior predictive distribution satisfies 
\begin{align} \label{esperanca}
p(\tilde{\bs Y} | \F_{n})  =\int_{\bs{\gamma}} p(\tilde{\bs Y} | \bs{\gamma}, \F_{n}) p(\bs{\gamma} | \F_{n}) d \bs{\gamma} ={\E}_{p(\bs{\gamma} | \F_{n})}\big(p(\tilde{\bs Y} | \bs{\gamma}, \F_{n})\big),
\end{align}
where $p(\bs{\gamma} | \F_{n})$ is the posterior distribution of \eqref{post} and $p(\tilde{\bs Y} |  \bs{\gamma},\F_{n})$ is the sampling density of the new data $\tilde{\bs Y}$, given parameters $\bs{\gamma}$ and $\F_{n}$. It can be computed iteratively by noting that
\begin{equation*}
p(\tilde{\bs Y} | \bs{\gamma}, \F_{n})=\prod_{k=1}^h p( Y_{n+k} | \bs{\gamma}, \F_{n+k-1}), 
\end{equation*}
where
\begin{equation*}
Y_{n+h}|\bs{\gamma},\F_{n+h-1}  \sim \mathrm{Beta}(\mu_{n+h},\nu), 
\end{equation*}
and
%
\begin{align}\label{beta}
\mu_t=g^{-1} \bigg( {\alpha} + \sum_{i=1}^{p}{\phi}_i \big( g(y_{t-i}) ) +\sum_{j=1}^{q}{\theta}_j {r}_{t-j} \bigg),
\end{align}
in which $r_t:=(g(Y_t)-g(\mu_t))$ is computed from $\mu_t$ and denotes an error term. It follows from \eqref{esperanca} that we can obtain samples from the posterior predictive distribution, if for each value of $\bs{\gamma^*}$ in a sample of the posterior distribution $p(\bs{\gamma}|\F_{n})$ we draw from  $p(\tilde{\bs Y} | \bs{\gamma},\F_{n})$. This is done by recursively computing $\mu_t$ given previous data and $\bs{\gamma}^*$, and then drawing a new value for $Y_{t}^*$ from the beta distribution in \eqref{beta}, for $t \in \{n+1, \cdots, n+h\}$.
%

Note that in deriving the posterior predictive distribution, for simplicity, we have considered the $\beta$ARMA without covariates. Including covariates in the model is straightforward as long as the new values $\bs{X}_{n+1},\cdots,\bs{X}_{n+h}$ are known. This is always possible if the covariates are deterministic functions, as in deterministic trends or seasonality. However, predicting future values for random covariates often requires the specification of a model to be feasible.
\subsection{Quasi-Unit roots testing} \label{qur}

Recall the following conditions, endemic to the study of ARMA models:
\begin{enumerate}
\item The AR and MA characteristic polynomials, given respectively by
\begin{equation}\label{chpol}\phi(z)=1-\phi_1z-\cdots-\phi_pz^p \quad \mbox{and}\quad \theta(z)=1+\theta_1z+\cdots+\theta_qz^q,\quad z\in\C,
\end{equation}
do not have common roots.
\item The AR characteristic polynomial satisfy $\phi(z)\neq0$ in the set $\Ca:=\{z\in\C:|z|\neq 1\}$.
\end{enumerate}
Condition 2 is a necessary and sufficient condition for the existence of a stationary solution for the ARMA equations and much attention is given to the presence of unit roots, that is, roots of the characteristic polynomial with absolute value equal to 1, which are usually associated with a random walk-like behavior.  However, model coefficients are estimated from a time series and are subject to uncertainty implying that exact unit roots occur with probability 0. Hence, in practice, we are concerned with the so-called quasi-unit roots, which are roots of the characteristic polynomial with a modulus close to 1. Quasi-unit roots also appear when a short-range dependence model such as the ARMA is applied to data presenting long-range dependence.

In the frequentist framework, unit roots are approached by using traditional tests, such as the Dickey-Fuller \citep{dickey1981}, Phillips-Perron \citep{phillipsperron}, among others. However, these tests are notorious for failing to detect unit roots in many contexts, especially in more complex dependence structures. Hence, a good practice in the frequentist approach is to calculate the roots of the estimated characteristic polynomial obtained by plugging in the estimated parameters, as a goodness-of-fit procedure. 

Technically, since observation-driven models are defined in a conditional fashion, there is no imposition of stationarity in the model. Nevertheless, since $g(\mu_t)=g(y_t)-r_t$, under conditions 1 and 2 the systematic component of a $\beta$ARMA model, namely \eqref{model}, can be written as
\begin{equation}\label{Tai4.7}
\phi(L)\big(g(Y_{t})-\bs x_t'\bs\beta-\omega\big)=\theta(L)r_t,  
\end{equation}
where $\omega:=\alpha/\phi(1)$, so that it satisfies the difference equations of an ARMA model with characteristic polynomials given in \eqref{chpol}. Hence, it is of interest to study unit roots in this context. However, testing a double bounded time series for unit roots is not even a well-posed problem since random walk behavior in a bounded domain is not well-defined \citep[but see][]{cav,cav2}. Besides, testing the original time series for unit roots is not the same as testing $g(\mu_t)$, since the latter depends on model parameters to be reconstructed. Hence, traditional unit root tests are useless in the context of $\beta$ARMA models. The only alternative under the frequentist framework is calculating the root of the approximated characteristic polynomial. But, there is no easy way to measure the uncertainty behind a given root of the approximated characteristic polynomial and no test to decide if a given quasi-unit root is or is not a consequence of a unit root in the underlying true model. The decision in this context ultimately falls under the user’s experience. 

Under the Bayesian framework, testing for the unit roots in the context of ARMA models has been explored in \cite{mar} and \cite{huerta}. The focus of these papers is on the specification of prior probabilities for the AR coefficients that facilitate testing for the presence of unit roots. In this work, we propose a simpler approach, that is more fitting to the context of GARMA-like models. Since the proposed Bayesian approach relies on obtaining samples for the joint posterior probability of the parameters, we can easily estimate the posterior probability that a unit root is present in the fitted model. This is attained by calculating the roots of the approximated characteristic polynomial for each sampled set of parameters in the posterior distribution and considering the proportion of roots whose modulus falls below a certain threshold. The finite sample performance of the proposed approach is studied in Section \ref{roots}.

\section{Monte Carlo Simulation Studies}  \label{simulation}
In this section, we present a Monte Carlo simulation study to assess the performance of the proposed approach in the context of $\beta$ARMA models. We perform three sets of simulations under a variety of scenarios to assess three different facets of the Bayesian analysis. In the first set, our interest lies in the accuracy of point estimation and the coverage of credible intervals. The second set analyzes the presence of quasi-unit roots in the model. Finally, we perform a sensitivity analysis on the prior parameter specification. 

The simulation was performed using \texttt{R} \citep{R}. Data generation was performed using  \texttt{R}  package \texttt{BTSR} \citep{BTSR}. A burn-in of 50 observations was applied in generating the time series, regardless of the sample size and parameters. Estimation was performed using the \texttt{RStan} package \citep{rstan}.  All data and codes are available at {\color{blue} \href{https://github.com/AlineFoersterGrande/Bayesian.BARMA}{github.com/AlineFoersterGrande/Bayesian.BARMA}}.

\subsection{Point and interval inference}\label{point}

In this section, we examine point and interval estimation of the proposed Bayesian approach in the context of $\beta$ARMA$(1,1)$ models. We simulate $\beta$ARMA$(1,1)$ model with $( \phi,\theta ) \in {\small\{(0.4, 0.4),(0.4, 0.6 ),(0.6, 0.4 ),(0.6, 0.6 ),(-0.5, -0.3 ),(-0.5, -0.4 ),(-0.3, -0.3 ),(-0.3,-0.4 )\}}$, $\nu\in\{50,100\}$, and sample sizes $n\in \{200,500\}$. We fit a $\beta$ARMA$(1,1)$  considering the following priors: $\nu\sim \mathrm{Gamma}(5,0.1)$, $\phi\sim N(\mu,\sigma^2)$ and $\theta\sim N(\mu,\sigma^2)$, where $\mu=0$ and $\sigma^2=20{,}000^2$. For each simulated data set, we ran two MCMC chains consisting of 2{,}000 samples each with a warm-up period of 50\%. Each experiment was replicated 50 times.  

\begin{table}[ht]
\scriptsize
\setlength{\tabcolsep}{5pt} 
\caption{Monte Carlo simulation results for $\beta$ARMA$(1,1)$ models. Presented are point and interval estimates obtained as the average posterior mean and average 95\% credible intervals.}\label{Table1} \vspace{0.2cm}
\begin{tabular}{ c|c|c|c|c|c|c|c|c} 
\hline
\multicolumn{3}{c}{DGP}& \multicolumn{3}{|c|}{$n=200$} &  \multicolumn{3}{c}{$n=500$} \\ 
\hline
\multirow{2}{*}{$\nu$} & \multirow{2}{*}{$\phi$} & \multirow{2}{*}{$\theta$} & $\nu$ & $\phi$ & $\theta$ & $\nu$ & $\phi$ & $\theta$ \\
&  &  & $CI_{0.95}(\nu)$ & $CI_{0.95}(\phi)$ & $CI_{0.95}(\theta)$ & $CI_{0.95}(\nu)$ & $CI_{0.95}(\phi)$ & $CI_{0.95}(\theta)$ \\ 
\hline
\multirow{16}{*}{50} 
& \multirow{2}{*}{0.40}  & \multirow{2}{*}{ 0.40} & 48.73 & 0.41 & 0.37 & 49.57 & 0.40 & 0.40 \\ 
&  &  & [39.98, 58.38] & [0.23, 0.58] & [0.18, 0.54] & [43.75, 55.79] & [0.28, 0.51] & [0.28, 0.51] \\ \cline{2-9}
& \multirow{2}{*}{ 0.40}  & \multirow{2}{*}{ 0.60} & 47.91 & 0.42 & 0.55 & 49.18 & 0.40 & 0.59\\ 
&  &  & [39.20, 57.47] & [0.26, 0.57] & [0.40, 0.68] & [43.35, 55.33] & [0.30, 0.50] & [0.50, 0.67] \\ \cline{2-9}
& \multirow{2}{*}{ 0.60}  & \multirow{2}{*}{ 0.40} & 48.28 & 0.60 & 0.38 & 49.27 & 0.60 & 0.40 \\ 
&  &  & [39.52, 57.80] & [0.46, 0.73] & [0.22, 0.53] & [43.43, 55.44] & [0.51, 0.68] & [0.30, 0.49] \\ \cline{2-9}
& \multirow{2}{*}{ 0.60}  & \multirow{2}{*}{ 0.60} & 47.55 & 0.60 & 0.57 & 49.01 & 0.60 & 0.59\\ 
&  &  & [38.98, 57.03] & [0.47, 0.72] & [0.43, 0.69] & [43.25, 55.15] & [0.52, 0.68] & [0.51, 0.67] \\ \cline{2-9}
& \multirow{2}{*}{ -0.50}  & \multirow{2}{*}{ -0.30} & 48.84 & -0.50 & -0.29 & 49.57 & -0.49 & -0.30 \\ 
&  &  & [40.08, 58.56] & [-0.66, -0.32] & [-0.46, -0.09] & [43.77, 55.75] & [-0.60, -0.39] & [-0.41, -0.18] \\ \cline{2-9}
& \multirow{2}{*}{ -0.50}  & \multirow{2}{*}{ -0.40} & 47.98 & -0.50 & -0.37 & 49.40 & -0.49 & -0.40\\ 
&  &  & [38.98, 58.03] & [-0.65, -0.34] & [-0.53, -0.18] & [43.60, 55.51] & [-0.59, -0.39] & [-0.50, -0.29] \\ \cline{2-9}
& \multirow{2}{*}{ -0.30}  & \multirow{2}{*}{ -0.30} & 49.12 & -0.31 & -0.27 & 49.77 & -0.30 & -0.30 \\ 
&  &  & [40.18, 58.89] & [-0.53, -0.08] & [-0.49, -0.03] & [43.94, 55.96] & [-0.44, -0.15] & [-0.44, -0.14] \\ \cline{2-9}
& \multirow{2}{*}{ -0.30}  & \multirow{2}{*}{ -0.40} & 49.03  & -0.32 & -0.36 & 49.72 & -0.30 & -0.39 \\ 
&  &  & [40.12, 58.75] & [-0.52, -0.11] & [-0.55, -0.16] & [43.90, 55.88] & [-0.43, -0.17] & [-0.51, -0.26] \\ 
\hline
\multirow{16}{*}{ 100} 
& \multirow{2}{*}{ 0.40}  & \multirow{2}{*}{ 0.40} & 93.11 & 0.42 & 0.36 & 97.23 & 0.40 & 0.39 \\ 
&  &  & [76.28, 111.53] & [0.24, 0.60] &  [0.16, 0.53] & [85.81, 109.40] & [0.29, 0.52] & [0.27, 0.50] \\ \cline{2-9}
& \multirow{2}{*}{ 0.40}  & \multirow{2}{*}{ 0.60} & 91.84 & 0.43 & 0.54 & 96.53 & 0.41 & 0.58 \\ 
&  &  & [75.20, 110.16] & [0.27, 0.58] & [0.38, 0.67] & [85.19, 108.60] & [0.31, 0.51] & [0.49, 0.66] \\ \cline{2-9}
& \multirow{2}{*}{ 0.60}  & \multirow{2}{*}{ 0.40} & 92.39 & 0.60 & 0.37 & 96.81 & 0.60 & 0.40\\ 
&  &  & [75.67, 110.73] & [0.46, 0.74] & [0.21, 0.52] & [85.40, 108.93] & [0.51, 0.68] & [0.29, 0.49] \\ \cline{2-9}
& \multirow{2}{*}{ 0.60}  & \multirow{2}{*}{ 0.60} & 90.83 & 0.61  & 0.55 & 96.07 & 0.60 & 0.58 \\ 
&  &  & [74.43, 108.78] & [0.48, 0.73] & [0.41, 0.67] & [84.74, 108.12] & [0.52, 0.68] & [0.50, 0.66] \\ \cline{2-9}
& \multirow{2}{*}{ -0.50}  & \multirow{2}{*}{ -0.30} & 91.45 & -0.50 & -0.25 & 97.42 & -0.50 & -0.30 \\ 
&  &  & [73.40, 111.82] & [-0.67, -0.33] & [-0.45, -0.04] & [85.88, 109.60] & [-0.60, -0.39] & [-0.41, -0.17] \\ \cline{2-9}
& \multirow{2}{*}{ -0.50}  & \multirow{2}{*}{ -0.40} & 91.82 & -0.51 & -0.35 & 97.07 & -0.50 & -0.39 \\ 
&  &  & [74.33, 111.39] & [-0.66, -0.35] & [-0.52, -0.16] & [85.63, 109.25] & [-0.60, -0.40] & [-0.49, -0.28] \\ \cline{2-9}
& \multirow{2}{*}{ -0.30}  & \multirow{2}{*}{ -0.30} & 93.68 & -0.32 & -0.26 & 97.74 & -0.30 & -0.29 \\ 
&  &  & [76.52, 112.28] & [-0.54, -0.08] & [-0.48, -0.02] & [86.19, 110.04] & [-0.45, -0.15] & [-0.44, -0.14] \\ \cline{2-9}
& \multirow{2}{*}{ -0.30}  & \multirow{2}{*}{ -0.40} & 93.52 & -0.33 & -0.35 & 97.67 & -0.31 & -0.39 \\ 
&  &  & [76.69, 111.98] & [-0.53, -0.12] & [-0.54, -0.14] & [86.18, 109.92] & [-0.43, -0.17] & [-0.51, -0.26] \\ 
\hline
\end{tabular}
\end{table}
The simulation results are shown in Table \ref{Table1}. For each parameter, the point estimates (average posterior mean) and average credible intervals (quantiles) are shown. From Table \ref{Table1}, we observe that the proposed Bayesian approach performs very well in all scenarios, with small biases for $\phi$ and $\theta$ even when $n=200$. It is worth noting that the specific values of $\phi$ and $\theta$ do not appear to have a significant effect on the credible intervals or point estimates for $\nu$.  Although $\nu$ has the highest variability, it has a smaller overall ratio between the credible interval length and the true parameter value.

Considering the point estimation for sample size $n=500$, the best scenario was $\nu=50$, $\phi=-0.3$ and $\theta=-0.3$, while the worst scenario was $\nu=100$ , $\phi=0.4$ and $\theta=0.6$. As expected, credible intervals in the $n=200$ scenario are considerably larger than those in the $n=500$ scenario. From the results, we conclude that the posterior estimates are close to the simulated values, which always fall within the mean credible interval.
\FloatBarrier
\subsection{Quasi-unit roots}  \label{roots}
In this section, we employ the posterior distribution to quantify the probability that the characteristic polynomial associated with \eqref{model} has a unit root. In this simulation we consider a model $\beta$ARMA$(2,0)$ with $\bs\phi=(\phi_1,\phi_2)$ given in the Table \ref{Table2}. The $\beta$ARMA$(2,0)$ model was fitted considering $n=500$, $\nu=100$ and the following priors: $\nu\sim \mathrm{Gamma}(5,0.1)$, $\phi_i\sim N(\mu,\sigma^2)$, where $\mu=0$ and $\sigma^2=20{,}000^2$ for $i\in\{1,2\}$. For each simulated series, two MCMC chains consisting of 2{,}000 samples each with a warm-up period of 50\% were run. The associated characteristic polynomial is given by $\phi(z)=1-\phi_1z-\phi_2z^2$ whose roots are $\frac{-\phi_1\pm\sqrt{\phi_1^2+4\phi_2}}{2\phi_2}$. The parameter values were chosen in order to provide characteristic polynomials with a variety of roots, some have the smallest modulus very close to 1 while others are far from 1. The smallest modulus of the roots for each parameter $\bs\phi$ is presented in Table \ref{Table2}'s third column. The results show the probability that the smallest modulus of the roots is smaller than $\{1.01,1.02,1.03,1.04,1.05\}$, where 1.05 is is frequently used as a practical rule-of-thumb to analyze unit-roots, in the sense that if the modulus is 1.05 or smaller, then we consider that a unit root is present in the true characteristic polynomial. 

Point estimates and credible intervals in this simulation were omitted to save up space, but the results were similar in quality to those presented in section \ref{point}. From the results, we observe that when the true modulus is $<1.05$, the posterior probability detects this with a very high probability, over 95\% in all cases. In the most difficult cases, when the true modulus is 1.05 and 1.057, the posterior probability that the modulus is $<1.05$ is still very high. As the true modulus moves away from 1.05, the posterior probability that the modulus is $<1.05$ decreases very rapidly to values close to 0.
\begin{table}[ht]
\setlength{\tabcolsep}{7pt} 
\renewcommand{\arraystretch}{1.1}
\caption{Posterior probability that the characteristic polynomial's roots have modulus smaller than a given threshold. The first two columns present the true parameter values with the smallest modulus of the characteristic polynomial's roots in the third column. Thresholds are presented in the last 5 columns.}\label{Table2} \vspace{0.2cm}
\centering
\begin{tabular}{r|r|c|c|c|c|c|c} 
 \hline  
\multicolumn{1}{c|}{$\phi_1$} & \multicolumn{1}{c|}{$\phi_2$}  & |root|  & $<1.01$ & $<1.02$ & $<1.03$ & $<1.04$ & $<1.05$ \\ 
 \hline
-0.25 & -0.95 & 1.026 & 0.050 & 0.463 & 0.929 & 0.992 & 0.996 \\   \hline
-0.15 & 0.80 & 1.028 & 0.034 & 0.184 & 0.504 & 0.816 & 0.957 \\   \hline
 0.20 & 0.75 & 1.029 & 0.081 & 0.356 & 0.743 & 0.946 & 0.995 \\ \hline
-0.30 & -0.90 & 1.050 & 0.002 & 0.016 & 0.133 & 0.474 & 0.823 \\  \hline
-0.10 & 0.80 & 1.057 & 0.030 & 0.164 & 0.483 & 0.809 & 0.954 \\  \hline
0.40 & 0.50 & 1.070 & 0.003 & 0.018 & 0.084 & 0.234 & 0.469 \\   \hline
0.10 & 0.80 & 1.100 & 0.004 & 0.005 & 0.015 & 0.057 & 0.159 \\   \hline
-0.90 & -0.60 & 1.300 & 0.000 & 0.000 & 0.000 & 0.000 & 0.000 \\   \hline
0.10 & -0.30 & 1.830 & 0.000 & 0.000 & 0.000 & 0.000 & 0.000 \\   \hline
0.30 & 0.10 & 2.000 & 0.002 & 0.002 & 0.002 & 0.002 & 0.002 \\   \hline
0.60 & -0.10 & 3.160 & 0.001 & 0.001 & 0.001 & 0.001 & 0.001 \\   \hline
\end{tabular}
\end{table}
\subsection{Sensitivity Analysis}  \label{sense}
In this section, we present a prior sensitivity analysis for parameter $\nu$, which presented the highest bias among the parameter estimates. The other parameters were well estimated using non-informative priors. For the sensitivity analysis, we consider $\beta$ARMA$(1,1)$ models with $\nu\in \{50,100\}$, $( \phi,\theta) \in \{(0.4, 0.6)$,  $(-0.3, -0.3)\}$ and sample size $n=200$. The \texttt{BTSR} package was used for data generation. For each simulated series, the posterior was estimated from two MCMC chains with $2{,}000$ samples each and a warm-up period of 50\%. 

The priors of $\phi$ and $\theta$ are the same as in Section \ref{point}, namely, $\phi\sim \mathrm{Normal}(\mu,\sigma^2)$ and $\theta\sim \mathrm{Normal}(\mu,\sigma^2)$ with $\mu=0$ and $\sigma^2=20{,}000^2$. As for $\nu$, we consider $\nu\sim\mathrm{Gamma}(\alpha,\beta)$ for a variety of $(\alpha,\beta)$ values, presented in Table \ref{alphasbetas} along with the respective induced prior mean and variance. Parameter choices induce priors that can be informative, as when $\nu=50$ and prior has mean 50 and variance 25, very non-informative, when variance is 2{,}000, and even very informative but equivocated, such as when the prior has mean 1 and variance 25, in which case, the prior probability that $\nu\leq 50$ is 0.9979 and that $\nu\leq 20$ is 0.9872. Each scenario was replicated 50 times. The results are presented in  Table \ref{Tabelasensitivity}, which shows the posterior mean and credible interval for each scenario, organized by the prior mean and the prior variance of $\nu$.

\begin{table}[ht]
\caption{Values of $(\alpha,\beta)$ used as prior for $\nu$ in the simulation and respective values of the prior mean and the prior variance.}\vspace{.3cm} \label{alphasbetas}
\centering
\begin{tabular}{ccc|c|c}
\cline{3-5}
 & & \multicolumn{3}{c}{Prior Mean}\\ \cline{3-5}
& &\multicolumn{1}{c|}{1} & \multicolumn{1}{c|}{50} & \multicolumn{1}{c}{100}\\ \hline
\multirow{3}{*}{\begin{tabular}{c}Prior\\Variance\end{tabular}} &\multicolumn{1}{|c|}{2000} & $(0.0005,0.0005)$ & $(1.25, 0.025)$ & $(5, 0.05)$\\ 
 &\multicolumn{1}{|c|}{500}  & $(0.002, 0.002)$ & $(5, 0.1)$ & $(20, 0.2)$\\ 
 &\multicolumn{1}{|c|}{25}   & $(0.04, 0.04)$ & $(100, 2)$ & $(400, 4)$\\ \hline
\end{tabular}
\end{table}
\begin{table}[ht]
\small
\setlength{\tabcolsep}{5pt} 
\caption{Results of the prior sensitivity analysis. The estimated values and credible intervals for each parameter are presented, organized by prior mean and variance.}\label{Tabelasensitivity} \vspace{0.2cm}
\begin{tabular}{ c|c|rc|rc|rc}
 \hline 
 \multirow{3}{*}{DGP} & \multirow{3}{*}{Prior Var}& \multicolumn{6}{c}{ Prior Mean}  \\
 \cline{3-8}
& & \multicolumn{2}{|c}{1} & \multicolumn{2}{|c}{50} & \multicolumn{2}{|c}{100} \\ 
\cline{3-8}
& & \multicolumn{1}{|c}{$\hat \nu$} & $CI_{0.95}(\nu)$& \multicolumn{1}{|c}{$\hat \nu$} & $CI_{0.95}(\nu)$& \multicolumn{1}{|c}{$\hat \nu$} & $CI_{0.95}(\nu)$\\
\hline
\multirow{9}{*}{\begin{tabular}{l} Scenario 1\\
$\nu=50$ \\ $\phi=0.40$ \\ $\theta=0.60$\end{tabular}} & \multirow{3}{*}{2000} &  47.85 & [39.00, 57.69] &  47.89 & [39.11, 57.63] &  49.06 &[40.19, 58.83]\\ 
&  & 0.42 & [0.26, 0.57] &  0.42 & [0.26, 0.57]  & 0.42 & [0.26, 0.57] \\ 
&  &  0.55 & [0.40, 0.68] & 0.55 & [0.40, 0.68]  & 0.55 & [0.40, 0.68] \\ 
 \cline{2-8}
& \multirow{3}{*}{500} &  47.80 &[38.92, 57.60]  & 47.91 & [39.20, 57.47]  &  52.31 & [43.52, 61.99] \\ 
&  &  0.42 & [0.26, 0.57] &  0.42 & [0.26, 0.57]  &  0.42 & [0.27, 0.56] \\ 
&  & 0.55 & [0.40, 0.68] & 0.55 & [0.40, 0.68]  & 0.56 & [0.41, 0.68] \\ 
 \cline{2-8}
& \multirow{3}{*}{25} & 46.97  & [38.31, 56.56]  & 48.81 & [42.28, 55.73]  & 81.90  & [74.90, 89.20] \\ 
&  &  0.42 & [0.26, 0.57] & 0.42 & [0.26, 0.57]  & 0.42  & [0.30, 0.54] \\ 
&  & 0.55 & [0.40, 0.68] & 0.55 & [0.41, 0.68]  & 0.56 & [0.45, 0.66] \\ 
 \hline
\multirow{9}{*}{\begin{tabular}{l} Scenario 2\\
$\nu=50$ \\ $\phi=-0.30$ \\ $\theta=-0.30$\end{tabular}} & \multirow{3}{*}{2000} &  49.12 & [40.07, 59.09] & 49.11 & [40.04, 59.05] & 50.31 & [41.15, 60.34]\\ 
&  & -0.31 & [-0.53, -0.08] & -0.32  & [-0.53, -0.08]  & -0.32 & [-0.53, -0.08] \\ 
&  &  -0.27 & [-0.49, -0.03] & -0.27 & [-0.49, -0.03]  & -0.27 & [-0.48, -0.03] \\ 
 \cline{2-8}
& \multirow{3}{*}{500} & 49.06 & [40.00, 58.95]  & 49.12 & [40.18, 58.89]  & 53.60 & [44.52, 63.61] \\ 
&  &  -0.32 & [-0.53, -0.08] &   -0.31 & [-0.53, -0.08]  &  -0.32 & [-0.53, -0.09] \\ 
&  & -0.27 & [-0.48, -0.03] &  -0.27 & [-0.49, -0.03]  &  -0.27 & [-0.48, -0.04] \\ 
 \cline{2-8}
& \multirow{3}{*}{25} & 48.22  & [39.26, 58.01]  & 49.45 & [42.85, 56.49]  & 82.62  & [75.59, 89.94] \\ 
&  & -0.31  & [-0.54, -0.08] & -0.32 & [-0.53, -0.08]  &  -0.32 & [-0.49, -0.14] \\ 
&  & -0.27 & [-0.49, -0.03] & -0.27 & [-0.48, -0.03]  & -0.27 & [-0.44, -0.09] \\ 
 \hline
\multirow{9}{*}{\begin{tabular}{l}  Scenario 3\\
$\nu=100$ \\ $\phi=0.40$ \\ $\theta=0.60$\end{tabular}} & \multirow{3}{*}{2000} &  95.84 & [78.10, 115.39] &  94.93 &  [77.41, 114.18] & 96.12  & [78.65, 115.25]\\ 
&  & 0.43 & [0.27, 0.58] & 0.43  &  [0.27, 0.58]  & 0.43 & [0.27, 0.58] \\ 
&  &  0.54 & [0.39, 0.67] & 0.54 & [0.39, 0.67]  & 0.54 & [0.39, 0.67] \\ 
 \cline{2-8}
& \multirow{3}{*}{500} & 95.79  &  [77.97, 115.45]  & 91.84 & [75.20, 110.16]  &  96.55 &  [80.10, 114,43] \\ 
&  & 0.43  & [0.27, 0.58] & 0.43  & [0.27, 0.58]  & 0.43  & [0.27, 0.58] \\ 
&  & 0.54 & [0.39, 0.67] & 0.54 & [0.38, 0.67]  & 0.54 & [0.39, 0.67] \\ 
 \cline{2-8}
& \multirow{3}{*}{25} & 92.34 & [75.16, 111.26]  & 65.65 & [56.90, 74.99]  & 99.04  & [90.55, 107.86] \\ 
&  & 0.43 & [0.27, 0.58] & 0.43 & [0.24, 0.60]  & 0.43  & [0.28, 0.57] \\ 
&  & 0.54 & [0.38, 0.67] & 0.53 & [0.35, 0.69]  & 0.54 & [0.39, 0.66] \\ 
 \hline
\multirow{9}{*}{\begin{tabular}{l} Scenario 4\\
$\nu=100$ \\ $\phi=-0.30$ \\ $\theta=-0.30$\end{tabular}} & \multirow{3}{*}{2000} & 97.99  & [79.70, 118.09] &  96.88 & [79.04, 116.54] & 98.06 & [80.35, 117.52]\\ 
&  & -0.32 & [-0.54, -0.09] &  -0.32 & [-0.54, -0.09]  & -0.32 & [-0.54, -0.09] \\ 
&  &  -0.26 &  [-0.47, -0.02] &  -0.26 &  [-0.47, -0.02]  & -0.26 &  [-0.47, -0.02] \\ 
 \cline{2-8}
& \multirow{3}{*}{500} & 97.83  & [79.68, 117.84]  & 93.68 & [76.52, 112.28]  & 98.28  & [81.55, 116.39] \\ 
&  & -0.32  & [-0.54, -0.09] &  -0.32 & [-0.54, -0.08]  & -0.32  & [-0.54, -0.09] \\ 
&  & -0.26 & [-0.48, -0.02] & -0.26 & [-0.48, -0.02]  & -0.26 & [-0.47, -0.02] \\ 
 \cline{2-8}
& \multirow{3}{*}{25} & 94.34 & [76.60, 113.72]  & 66.15 & [57.32, 75.64]  & 99.48  & [90.98, 108.41] \\ 
&  & -0.32  & [-0.54, -0.08] & -0.32 & [-0.58, -0.04]  & -0.32  & [-0.54, -0.09] \\ 
&  & -0.26 & [-0.48, -0.02] & -0.26 & [-0.51, 0.03]  & -0.26 & [-0.47, -0.03] \\ 
 \hline
\end{tabular}
\end{table}
\FloatBarrier
The changes in the prior for $\nu$ did not affect point estimates for $\phi$ and $\theta$. Overall, point estimates and credible intervals for $\nu$ are on par with the ones presented in Section \ref{point}, with a few exceptions. In scenarios 1 and 2, where the true value is $\nu=50$, when the prior mean is 100 and the prior variance is 25, point estimates are very biased, and credible intervals do not contain 50. In this case, the prior probability that $\nu<85$ is only 0.008, which flattens out the posterior in the vicinity of the true value of $\nu$, 50, where the likelihood is likely to be higher. In scenarios 3 and 4, for which $\nu=100$, when the prior variance is 25 and the prior mean is 50, essentially the same happens with 0.9972 prior probability that $\nu\leq 65$.

Interestingly, the most extreme case where the prior mean is 1 and the prior variance is 25 yielded good overall results, despite the very small prior probability for values of $\nu$ distant from 1. This is so because 1 is very distant from the likelihood peaks, hence the likelihood values in the vicinity of 1 are very small, counteracting the effect of the prior. Near the likelihood peaks, the prior is basically flat so the prior probability is essentially dominated by the likelihood, allowing for the identification of the peaks in the posterior probability. 

Comparing the results presented in Table \ref{Tabelasensitivity} and Table \ref{Table1}, we observe that applying a more informative prior, for which the mean matches the data-generating process, yielded slightly better results. The best results were obtained when the prior mean is 100 and the worst when the prior variance is 25. We conclude that, as expected, the parameters are well estimated when the prior is uninformative, but $\nu$ is sensible to poor choices of informative priors, which should in practice be avoided. 
%
%
\section{Empirical Application}  \label{aplication}
\begin{figure}[ht] 
\begin{center} 
\includegraphics[width=0.65\textwidth]{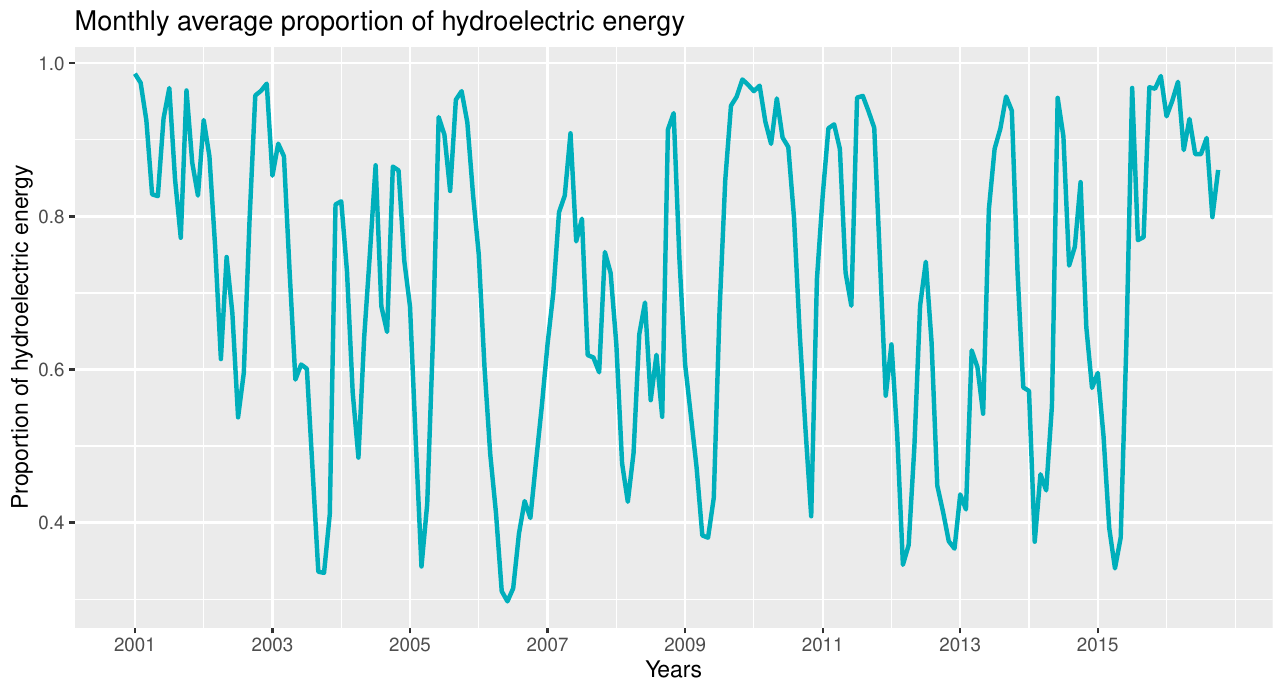}
\caption{Monthly average proportion of hydroelectric energy stored in southern Brazil.} \label{fig0}
\end{center}
\end{figure}

In this section, we present an application of the proposed methodology to a real data set. We consider the monthly average proportion of hydroelectric energy stored in southern Brazil. The time series consists of monthly averages ranging from January 2001 to October 2016, yielding a sample size of $n=190$ observations (Figure \ref{fig0}). Data from November 2016 to April 2017 ($n=6$), was reserved for forecasting purposes. The same time series was used in \cite{Scher}, 
where the authors fit 5 different models to the data and present a comparison among them in terms of out-of-sample forecast. The best predictive model was a $\beta$ARMA$(1,1)$, with parameters $\alpha=0.3452$, $\nu=11.7593$, $\phi= 0.5235$ and $\theta=0.3588$, obtained using the conditional maximum likelihood approach of \cite{Rocha2009}.  In this section we focus on applying the proposed Bayesian approach to compare the results with those obtained by \cite{Scher}.

\begin{figure}[ht] 
\begin{center} 
\includegraphics[width=0.5\textwidth]{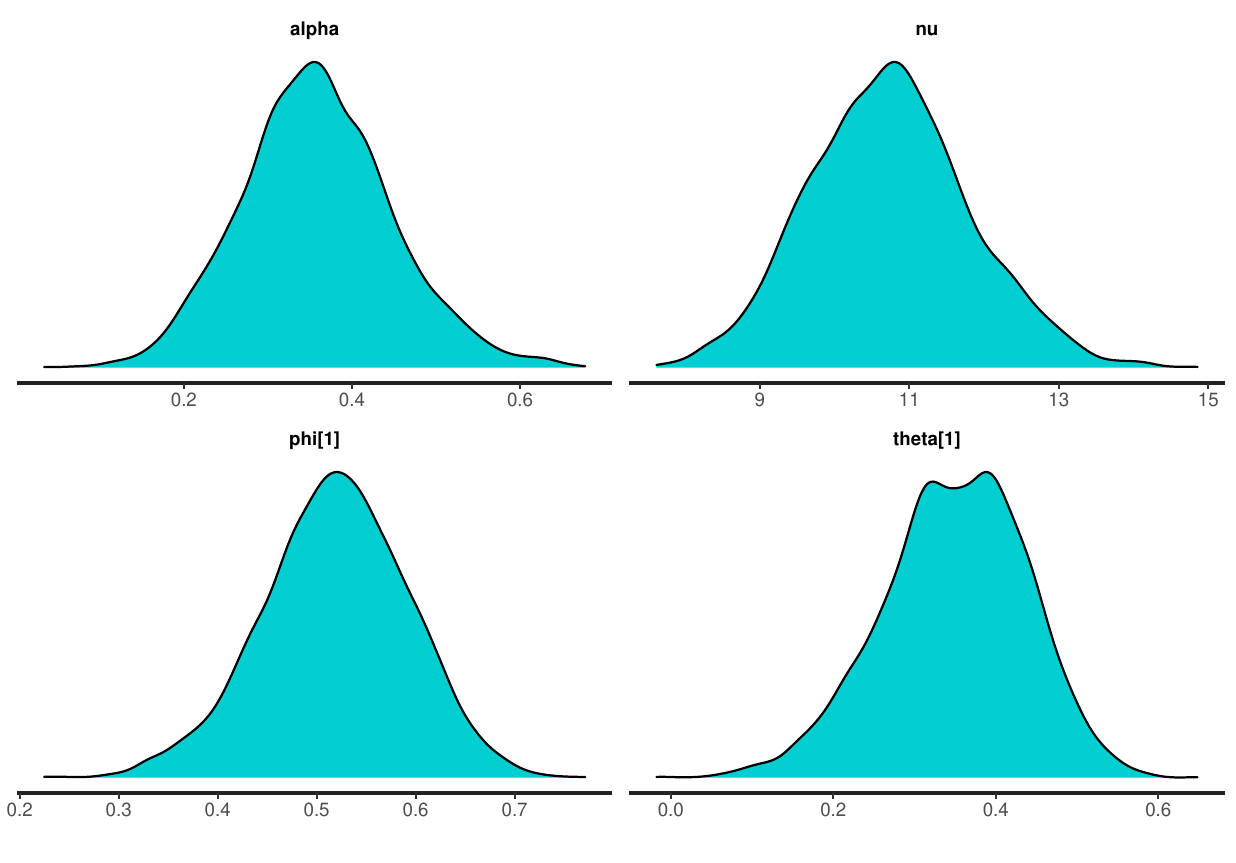}
\caption{Posterior density plots of parameters $\alpha$ (top left), $\nu$ (top right), $\phi$ (bottom left) and $\theta$ (bottom right).} \label{fig1}
\end{center}
\end{figure}

\begin{figure}[ht] 
\begin{center} 
\includegraphics[width=0.5\textwidth]{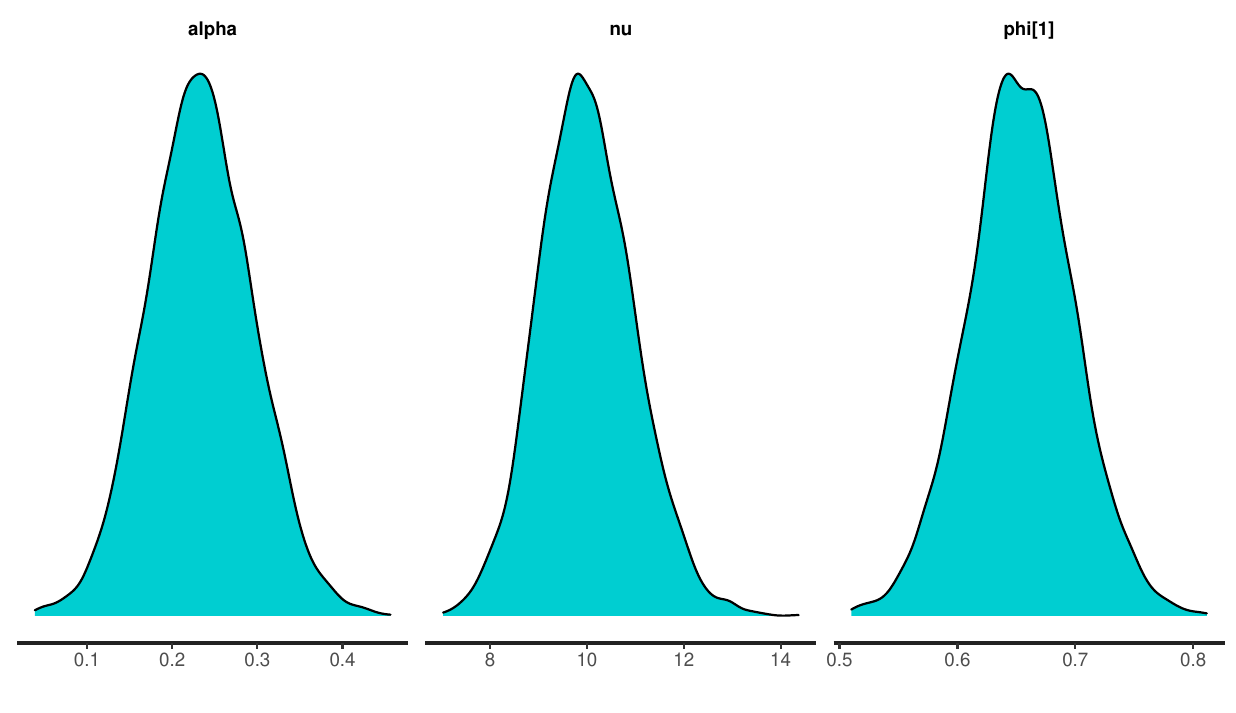}
\caption{Posterior density plots of parameters $\alpha$ (left), $\nu$ (middle) and $\phi$ (right).} \label{fig2}
\end{center}
\end{figure}

Under the proposed Bayesian approach we fit a $\beta$ARMA$(1,1)$ and a $\beta$ARMA$(1,0)$ considering the following priors: $\nu\sim \mathrm{Gamma}(5,0.1)$, $\alpha\sim\mathrm{Uniform}(-1, 1)$ (intercept), $\phi\sim N(\mu,\sigma^2)$ and $\theta\sim N(\mu,\sigma^2)$, where $\mu=0$ and $\sigma^2=20{,}000^2$. For each model, we run four chains consisting of 2{,}000 samples each with a warm-up period of 50\%. In Table \ref{TableSummaryApplication} we compile the results, presenting point estimates (posterior mean), 95\% credible interval, and effective sample size (ESS). 
%
%
\begin{table}[ht]
\setlength{\tabcolsep}{5pt}
\caption{Fitted $\beta$ARMA$(1,1)$ and $\beta$ARMA$(1,0)$ models for the proportion of hydroelectric energy stored in southern Brazil.}\label{TableSummaryApplication} \vspace{0.2cm}
\centering
\begin{tabular}{ c|c|c|c|c} 
\hline 
Parameter&$\alpha$ & $\nu$ & $\phi$ & $\theta$ \\
\hline
Estimates&0.36 & 10.73 & 0.52 & 0.35 \\ 
$\mathrm{CI}_{0.95}$ & $[0.19, 0.55]$ & $[8.71, 12.94]$ & $[0.37, 0.66]$  & $[0.16, 0.51]$  \\ 
ESS & 2094 & 3517 & 2021 & 2330 \\ 
\hline
Estimates & 0.23 & 10.01 & 0.65 & $-$  \\ 
$\mathrm{CI}_{0.95}$ & $[0.12, 0.36]$ & $[8.13, 12.02]$ & $[0.57, 0.75]$  & $-$  \\ 
ESS & 2245 & 2887 & 2368 & $-$  \\ 
\hline
\end{tabular}
\end{table}
From Table \ref{TableSummaryApplication}, point estimates obtained with the proposed approach are close to the ones obtained in \cite{Scher}. In both models, none of the credible intervals contain 0, which bears evidence for the relevance of including these parameters in the model. The ESS of all parameters is high, above 2{,}000, indicating a very low correlation among the sampled values from the posterior, as expected from the HMC approach. 
To provide additional information about the posterior distributions of the model parameters, we present in Figures \ref{fig1} and \ref{fig2} the density plots for each parameter.

 A Bayes Factor model selection approach was employed to define which model to consider in this application. The idea is to compare several models in increasing order of complexity using a Bayes Factor approach to perform model selection. This is somewhat similar to the model selection of stationary ARMA models in the Box and Jenkins framework. Table \ref{MarginalLikelihoodApplication} presents the values of the (log) marginal likelihoods for each model. Log-Bayes Factor values, which measure how much more likely the data are under the null compared to the alternative model, are obtained as the difference in the log-marginal likelihood between the null and the alternative model and can be computed directly from the table. 

%
%
%
\FloatBarrier


 We note that the $\beta$ARMA$(1,0)$ model, with the largest log marginal likelihood, presents a log Bayes Factor of 6.19 when compared to the $\beta$ARMA$(1,1)$ with the second largest log marginal likelihood. This indicates decisive evidence in favor of the  $\beta$ARMA$(1,0)$ model and motivates its inclusion in this analysis. Furthermore, since the $\beta$ARMA$(1,1)$ performs well when compared to all other models considered (with a log Bayes Factor to the next best model of 4.46), and is of the same order used in \cite{Scher}, allowing for a more direct comparison, it was also included.

We employ both models to generate out-of-sample forecasts and compare them to the predictions obtained for the $\beta$ARMA$(1,1)$ in \cite{Scher}. As previously mentioned, the last six months (November 2016 to April 2017) were reserved for this purpose. \cite{Scher} only reports the mean absolute error (MAE) of forecast in their work, hence, we report the performance of the $h$-steps ahead forecasts obtained with the predictive posterior approach presented in Section \ref{pred} in terms of MAE. The results are presented in Table \ref{TablePredictionApplication}.
\begin{table}[ht]
\setlength{\tabcolsep}{9pt}
\caption{Log Marginal Likelihood Results.}\label{MarginalLikelihoodApplication} \vspace{0.2cm}
\centering
\begin{tabular}{c|c|c|c|c|c}
\hline
$\beta$ARMA & $(0,1)$ & $(1,0)$ & $(1,1)$ & $(1,2)$ & $(2,1)$\\ \hline
log-ML      & 107.26 &  117.91  &  111.72  &  99.69  & 101.47\\ \hline
\end{tabular}
\end{table}
\begin{table}[H]
\setlength{\tabcolsep}{5pt}
\caption{Results from the $h$-steps ahead forecasts for the Bayesian and frequentist approach. Presented are the MAE of forecasting for each model.}\label{TablePredictionApplication} \vspace{0.2cm}
\centering
\begin{tabular}{ c|c|c|c|c|c|c} 
\hline  
\multicolumn{1}{c|}{Model} &  $h=1$ &  $h=2$ &  $h=3$ &  $h=4$ &  $h=5$ &  $h=6$  \\
\hline 
Bayesian $\beta$ARMA$(1,1)$ & 0.1184 & 0.1414 & 0.1371 & 0.1537 & 0.1778 & 0.1949\\
\hline 
Bayesian $\beta$ARMA$(1,0)$ & 0.0964 & 0.1331 & 0.1372 & 0.1572 & 0.1839 & 0.2020\\
\hline
Frequentist $\beta$ARMA$(1,1)$ & 0.1244 & 0.1444 & 0.1364 & 0.1484 &  0.1694 & 0.1839  \\
\hline  
\end{tabular} 
\end{table}
The results in Table \ref{TablePredictionApplication} show that for $h\in\{1,2\}$, the Bayesian $\beta$ARMA$(1,0)$ presented the best-forecasted values in terms of MAE, followed by the Bayesian $\beta$ARMA$(1,1)$. For $h\in\{3,4,5,6\}$, the frequentist $\beta$ARMA$(1,1)$ presented the best forecasted values. Comparing the Bayesian models, $\beta$ARMA$(1,0)$ is uniformly outperformed by the $\beta$ARMA$(1,1)$ for $h\in\{3,4,5,6\}$, and only performs better for $h\in\{1,2\}$, as mentioned before.
\section{Discussion}
In this paper, we studied parameter estimation of $\beta$ARMA models under the Bayesian framework. The model considered standard forms for the prior distribution while the posterior distribution of the parameters was obtained through the Hamiltonian Markov Chain Monte Carlo. We proposed the use of posterior probabilities to approach the problem of unit roots in the model's systematic component. We also considered a procedure consisting of a series of Bayes Factor tests to perform model selection and addressed the problem of forecasting using the posterior predictive distribution.

Monte Carlo simulations were conducted to evaluate the model's performance, which was divided into three parts. The first, presented in Section \ref{point}, provided evidence that the proposed Bayesian approach for $\beta$ARMA$(1,1)$ models performs well in terms of point and interval estimation. The method was able to accurately estimate model parameters in a variety of scenarios, including different values of $\nu$, $\phi$, $\theta$, and sample sizes. Even with a small sample size, the results showed that point estimation had only small biases for the parameters. The credible intervals were found to generally contain the true value, indicating good precision in the parameter estimates.

In Section \ref{roots}, we put forth a Monte Carlo simulation to assess the proposed approach to unit roots in the context of $\beta$ARMA$(2,0)$ models. A variety of parameters were chosen to provide characteristic polynomials whose roots are near and far from one in modulus. The results show that the posterior probability can be successfully employed to quantify the uncertainty associated with a given quasi-unit root with high precision. Overall, the proposed Bayesian approach provides a reliable way of detecting unit roots in $\beta$ARMA models.

In Section \ref{sense}, we performed a prior sensitivity analysis for parameter $\nu$. We simulated a variety of $\beta$ARMA$(1,1)$ models, with different combinations of parameters, different sample sizes, and different priors for $\nu$.  The results showed that the Bayesian approach performs very well when the prior is non-informative. When the prior was informative and the information provided by the prior was correctly specified (low variance and correctly specified mean), the results showed a small improvement in the estimated values. However, in certain cases where the prior specification was informative, but the information provided was incorrect (low variance but misspecified mean, far from the true parameter), the effects in the posterior were evident, resulting in a high bias in point estimates and distortions in the credible intervals. The best estimates were obtained when the prior mean was 100, while the worst were obtained when the prior variance was 25. This suggests that caution should be used when specifying priors for $\nu$, particularly when the prior variance is low.

In Section \ref{aplication}, we presented a real data application of the proposed Bayesian approach to the monthly average proportion of hydroelectric energy stored in southern Brazil. Two models, $\beta$ARMA$(1,1)$ and $\beta$ARMA$(1,0)$, were fitted using the Bayesian approach. We found that the $\beta$ARMA$(1,0)$ model is selected by the log Bayes factor approach, followed by a $\beta$ARMA$(1,1)$. Both models were compared in terms of their out-of-sample forecasting capabilities to a $\beta$ARMA$(1,1)$ selected with a frequentist approach in \cite{Scher}.  For most forecasting horizons, the Bayesian $\beta$ARMA$(1,1)$ model generated more accurate forecasts than the $\beta$ARMA$(1,0)$. Both Bayesian models outperform the frequentist model in the first few steps.

In summary, the proposed Bayesian approach provides a competitive, reliable, and flexible tool for the estimation of $\beta$ARMA models, with the potential to contribute to the dissemination of $\beta$ARMA models and Bayesian analysis. One interesting topic for future work is the development of an \texttt{R} package (or a user-friendly script) capable of fitting the proposed Bayesian approach based on the Hamiltonian MCMC to further promote the method.
\small
\subsection*{Acknowledgments}
Aline F. Grande was supported by the Coordena\c c\~ao de Aperfeiçoamento de Pessoal de N\'ivel Superior – Brasil (CAPES). The authors are also grateful to Professor Dani Gamerman for his valuable comments and suggestions.
\subsection*{Disclosure Statement}
The authors report there are no competing interests to declare.
\normalsize

\bibliographystyle{apalike}
\bibliography{exemplo}

\end{document}